\documentclass[10pt,a4paper]{article}

\usepackage{latexsym}
\usepackage{amssymb}

\newif\ifpdf
\ifx\pdfoutput\undefined
   \pdffalse
\else
   \pdfoutput=1
   \pdftrue
\fi

\ifpdf
   \usepackage[pdftex,
               pdftitle={Portfolio Optimization with Spectral Measures of Risk},
               pdfsubject={Portfolio Optimization with Spectral Measures of Risk},
               pdfkeywords={Expected Shortfall; Risk measure; worst conditional expectation; tail conditional expectation;
               value-at-risk (VaR); conditional value-at-risk (CVaR); tail mean; coherence; quantile;
               sub-additivity},
               pdfauthor={},
               pdfstartview=FitBH,
               pdfview=FitBH,
               colorlinks=true]{hyperref}
\else
\fi

\newtheorem{theorem}{Theorem}[section]

\newtheorem{remark}[theorem]{Remark}
\newtheorem{example}[theorem]{Example}
\newtheorem{proposition}[theorem]{Proposition}
\newtheorem{definition}[theorem]{Definition}

\title{Portfolio Optimization with Spectral Measures of Risk}
\author{C. Acerbi and P. Simonetti \\ \\
\it \small Abaxbank, Corso Monforte 34, 20122 Milano (Italy) }

\begin{document}
\maketitle

\begin{abstract} 
We study Spectral Measures of Risk from the perspective of portfolio optimization. We derive exact results which extend to general Spectral Measures $M_\phi$ the Pflug--Rockafellar--Uryasev  methodology for the minimization of $\alpha$--Expected Shortfall. The minimization problem of a spectral measure is shown to be equivalent to the minimization of a suitable function which contains additional parameters, but displays analytical properties (piecewise linearity and convexity in all arguments, absence of sorting subroutines) which allow for efficient minimization procedures. 

In doing so we also reveal a new picture where the classical risk--reward problem \`a la Markowitz (minimizing risks with constrained returns or maximizing returns with constrained risks) is shown to coincide to the unconstrained optimization of a single suitable spectral measure. In other words, minimizing a spectral measure turns out to be  already an optimization process itself, where risk minimization and returns maximization cannot be disentangled from each other.

\end{abstract}

\section{Introduction}

Spectral Measures of Risk  \cite{A01} represent the class of all Coherent Measures of Risk  \cite{ADEH97,ADEH99} which display  the additional properties of law--invariance and comonotonic additivity \cite{Kus01,Ta02}. The requisite of law--invariance, being a necessary condition for a statistics to be estimable from empirical data, ensures that these  measures are suitable for industrial applications. Consistent estimators of Spectral Measures have indeed been given in \cite{A01} via scenario--based explicit representations. 

This class of statistics also  provides  a connection between risk measures and subjective risk aversion. Every spectral measure $M_\phi$ is in fact characterized by a particular ``risk aversion function'' $\phi$: the space of all possible coherent spectral measures is spanned by  all the possible risk aversions $\phi$ of a rational investor. In other words, the measure of risk $M_\phi$ encodes via $\phi$ the subjective risk attitude of a particular investor and it is coherent if and only if the attitude of the investor is  really risk averse (Theorem 4.1 of \cite{A01}).

In this paper we study Spectral Measures in the context of portfolio optimization. The fact that coherent measures are convex in the space of portfolios makes them excellent candidate for alternative risk indicators in the Markowitz plane in place of the traditional standard deviation. It is natural in other words to study the classical constrained optimization problem where $M_\phi$ is minimized for a given level of expected return or vice versa expected return is maximized for a given level of risk $M_\phi$.

The problem may however seem easier than it actually is. Despite the convexity property, which ensures the existence and unicity of a global minimum in the risk surface, Spectral Measures are difficult statistics to minimize unless specific tricky procedures are adopted. The problem hidden in their minimization is correctly perceived if one remembers that their estimators always contain a sorting procedure of portfolio scenarios (see the presence of ordered statistics in eq. (\ref{estM}) below). The dependence of estimators on portfolio parameters is not explicitly analytic because smooth changes in the parameters, induce discrete permutations of scenarios. Ordinary minimization techniques fail because the target function contains a {\tt SORT(scenarios)} subroutine. 

The problem was elegantly solved for the particular case of Expected Shortfall ($ES$) (or Conditional Value at Risk, $CVaR$) by Pflug--Rockafellar--Uryasev \cite{Pf00,RU00,RU01} who showed that the minimization of $ES$ is equivalent to the minimization of another function which depends on an auxiliary variable but is free from any sorting procedure. The mathematical tractability of this function allows for highly efficient procedures.

The first goal of this paper is to generalize the results of \cite{Pf00,RU00,RU01} to any spectral measure $M_\phi$. In doing so, however, we discover that the minimization of a Spectral Measure, is in some sense already a risk minimization and a return maximization at the same time. A new picture emerges where the two processes appear to be indistinguishable. Constrained risk--reward optimization is in fact shown to coincide with unconstrained minimization of particular spectral measures which interpolate between risks and rewards. 

The efficient frontier of a classical Markowitz $ES$--vs--return problem turns out to be only the simplest example of a wide range of optimization problems where the preference among portfolios is driven by the risk aversion function $\phi$. A new picture emerges where new  optimization problems can be solved, which have no analogous in a traditional two dimensional risk--reward Markowitz plane.

\subsection{Notation and definitions}

We will denote by $\Pi= \Pi(\vec{w})$  a portfolio or its value alternatively and suppose that it depends on a set of parameters or {\em``weights''} $w_k$ $k=1, \ldots, W$. We denote with $X(\vec{w})=\Delta\Pi(\vec{w})$ the {\em profit and loss} (p\&l) r.v. of the portfolio on a chosen time horizon $\Delta t$.


Following  \cite{A01}  we define\footnote{We adopt the following definitions for the generalized inverse probability distribution functions
\begin{equation} \nonumber
\begin{array}{l} 
F^\leftarrow_X (\alpha) \equiv  x_{\alpha} \equiv \inf\{x | F_X(x)\geq \alpha \}  \\
F^\rightarrow_X (\alpha) \equiv  x^{\alpha} \equiv \inf\{x | F_X(x)> \alpha \} 
\end{array} \nonumber
\end{equation}
Where they coincide, they define $F^{-1}(\alpha)$.  As indicated above this is also the definition  of lower and upper quantile $x_{\alpha}$ and $x^{\alpha}$.
}: 

\begin{definition}[Spectral Measure of Risk] a Spectral Measure of Risk is any measure of the kind
\begin{equation} \label{Mphi}
M_\phi (X) = - \int_0^1 \; \phi(p) F_X^\leftarrow(p)\, dp
\end{equation}
where $\phi$ is a real function on the interval $[0,1]$.
\end{definition}

\begin{definition}[Admissible Risk Aversion Function] \label{ARAF} a (generalized) function $\phi$ on the interval $[0,1]$ is said to be an Admissible Risk Aversion Function if it is of the kind
\begin{equation}
\phi(p)= c\, \delta(p) + \tilde{\phi}(p)
\end{equation}
where $\delta$ is a Dirac delta--distribution, $c\in [0,1]$, and $\tilde{\phi}:[0,1]\rightarrow \mathrm{R}$ satisfies
\begin{enumerate}
\item $\tilde{\phi}(p)\geq 0 \:\:\: \forall p$
\item $p_1 < p_2 \; \Rightarrow \; \tilde{\phi}(p_1)\geq \tilde{\phi}(p_2) $
\item $\int_0^1 \,\tilde{\phi}(p) \,dp =1-c$
\end{enumerate}
An Admissible Risk Aversion Function is said to be non--singular if $c=0$.
\end{definition}

A Spectral Measure of Risk is a Coherent Measure of Risk in the sense of \cite{ADEH99} if and only if $\phi$ is an admissible risk spectrum \cite{A01}. 

The possible singular part $c\,\delta(p)$ in the risk aversion function brings into the Spectral Measure a certain amount of ``worst case scenarios'' $F^\leftarrow_X(0)= {\rm ess}\inf\{X\}$. For instance, as a limiting case, one can study the  measure defined by a $\phi$ which has $c=1$ and $\tilde{\phi}=0$:
\begin{equation}
M_{worst}(X) \equiv ES_{\alpha=0}(X) = -F^\leftarrow_X(0) = -{\rm ess}\inf\{X\}
\end{equation}
This is a perfectly legitimate coherent measure of risk which literally represents the worst case scenario of the probability distribution of $X$. It has finite value only if the distribution of $X$ is bounded from below, otherwise it blows to $+\infty$.

\section{Minimizing the Expected Shortfall: the Pflug--Rockafellar--Uryasev method}

The problem of the optimization of the Expected Shortfall has been widely investigated by Pflug, Rockafellar and Uryasev \cite{Pf00,RU00,RU01}. In this section we review their results both for notational convenience and for getting some insight that will help us generalizing these results to the case of a general Spectral Measure\footnote{The authors of \cite{RU00,RU01} prefer the terminology Conditional Value at Risk (CVaR) for what we call Expected Shortfall (ES). }. 

The fundamental result of  \cite{RU00,RU01} is that the problem of minimization of the ES 
\begin{equation}
ES_{\alpha} (X(\vec{w})) = -\frac{1}{\alpha} \; \int_0^\alpha \, F^\leftarrow_{X(\vec{w})}(p) \, dp
\end{equation}
is shown to be equivalent to the problem of minimization of the following function containing an auxiliary variable $\psi$:
\begin{equation} \label{gammaURcont}
\Gamma_{\alpha}(X(\vec{w}),\psi) = -\psi + \frac{1}{\alpha}\mathrm{E}[-X(\vec{w})+\psi]^+
\end{equation}
More explicitly, the minimum achieved by $ES_{\alpha} (X(\vec{w}))$ by varying $\vec{w}$ equals the minimum achieved by $\Gamma_{\alpha}(X(\vec{w}),\psi)$ by varying both $\vec{w}$ and $\psi$
\begin{equation}
\min_{\vec{w}} ES_{\alpha} (X(\vec{w})) = \min_{\vec{w},\psi}\Gamma_{\alpha}(X(\vec{w}),\psi)
\end{equation}
As a byproduct of the minimization, the auxiliary variable $\psi$ in the minimum of $\Gamma_{\alpha}$ can be shown to assume a value in the interval defined by the upper and lower $\alpha$--quantiles
 $[x_\alpha,x^\alpha]$.
\begin{equation} \label{argmin}
\min_\psi \Gamma_{\alpha}(X,\psi) =  \Gamma_{\alpha}(X,\psi^*) = ES_{\alpha} (X) \;\;\;\;\;\;\;\;\; \forall \psi^* \in[x_\alpha,x^\alpha]
\end{equation}
When the upper and lower quantiles coincide, the argument $\psi$ in the minimum equals {\em the quantile} ($\psi^*=x_\alpha=x^\alpha$).

To check eq. (\ref{argmin}) we rewrite
\begin{eqnarray}
\Gamma_{\alpha}(X,\psi) &=& -\psi + \frac{1}{\alpha}\mathrm{E}[-X+\psi]^+  \\
&=& -\psi + \frac{1}{\alpha} \int dP_X (-X+\psi) \mathrm{1}_{\{X\leq \psi\}}\nonumber\\
&=& -\frac{1}{\alpha} \mathrm{E}[X \, \mathrm{1}_{\{X\leq \psi\}}] - \psi \left( 1-\frac{ F_X(\psi)}{\alpha} \right)\nonumber
\end{eqnarray}
This expression is well known to be invariant if evaluated in any $\psi \in[x_\alpha,x^\alpha]$ and its value there coincides with $ES_\alpha(X)$ (see \cite{AT01}).

To show that $\Gamma_{\alpha}(X,\psi)$ reaches its minimum in $\psi\in[x_\alpha,x^\alpha]$ we compute
\begin{eqnarray}
\frac{\partial \Gamma_{\alpha}(X,\psi)}{\partial \psi} &=& \frac{\partial}{\partial \psi} 
\left( 
-\psi + \frac{1}{\alpha} \int dP_X (-X+\psi)^+ 
\right) \\
&=& -1 + \frac{1}{\alpha} \int dP_X  \mathrm{1}_{\{X\leq \psi\}} \nonumber\\
&=& -1 + \frac{F_X(\psi)}{\alpha} \nonumber
\end{eqnarray}
from which one immediately obtains 
\begin{eqnarray}
\frac{\partial \Gamma_{\alpha}(X,\psi)}{\partial \psi} >0 & if & \psi> x^\alpha \\
\frac{\partial \Gamma_{\alpha}(X,\psi)}{\partial \psi} <0 & if & \psi< x_\alpha \nonumber
\end{eqnarray}
which ensures that the minimum is reached in $\psi\in[x_\alpha,x^\alpha]$.

The unfamiliar reader could be led to think that the presence of an auxiliary variable $\psi$ makes $\Gamma_{\alpha}(X(\vec{w}),\psi)$ more difficult to minimize than $ES_{\alpha} (X(\vec{w}))$. On the contrary, in any concrete case,  minimizing  $\Gamma_{\alpha}(X(\vec{w}),\psi)$ is extraordinarily simpler than minimizing $ES_{\alpha} (X(\vec{w}))$ despite the presence of the additional variable. To convince ourselves, let us study the discrete case where the probability distribution of $X$ is accessible only via $N$ empirical  realizations $X_i=X_i(\vec{w})$ of the portfolio p\&l. In other words, we want to minimize the following estimator\footnote{We adopt standard definition for the ordered statistics $X_{i:N}$: $\{X_{i:N}| i=1,\ldots,N\}=\{X_{i}| i=1,\ldots,N\}$ as sets and $X_{i:N}\leq X_{i+1:N}\:\forall i$.} of $ES$ (see \cite{AT2})
\begin{equation} \label{estES}
ES_\alpha^{(N)}(X(\vec{w})) = - \frac{1}{[ N\alpha ]} \sum_{i=1}^{[ N\alpha ]} X_{i:N}(\vec{w})
\end{equation}
The result of Pflug, Rockafellar and Uryasev teaches us that we can alternatively minimize 
the discrete version of (\ref{gammaURcont}) given by 
\begin{equation}\label{gammaURdiscrete}
\Gamma_{\alpha}^{(N)}(X(\vec{w}),\psi) = -\psi + \frac{1}{[N\alpha]} \sum_{i=1}^N (\psi-X_i(\vec{w}))^+ 
\end{equation}
The reader should give a careful glance at the two expressions above: in eq. (\ref{estES}), the presence of the ``ordered statistics'' $X_{i:N}$ signals a severe problem hidden in the dependence of $ES_\alpha^{(N)}$ on $\vec{w}$. Varying the parameters $\vec{w}$, the ordered statistics are sorted in a discrete way so that the dependence from the parameters is not explicitly analytic, with the consequence that standard minimization techniques fail.

In eq. (\ref{gammaURdiscrete}) on the contrary, the ordered statistics disappear and the dependence of $\Gamma_{\alpha}^{(N)}$ is manifestly analytic in its arguments $\psi$ and $X_i$. More exactly  $\Gamma_{\alpha}^{(N)}$ is manifestly {\em piecewise linear} and can be shown to be {\em convex} in all $\psi$ and $X_i$.
If $X_i(\vec{w})$ is in turn linear in $\vec{w}$, as in the case when the parameters $w_k$ are the weights of $W$ assets in the portfolio, the properties of piecewise linearity and convexity extend to the dependence of $\Gamma_{\alpha}^{(N)}$ on $\vec{w}$.

We now try to understand  what is {\em the role} of the auxiliary variable $\psi$. To do this let us  derive the  extremal conditions of $\Gamma_{\alpha}^{(N)}$:
\begin{eqnarray}
0 	&=& \frac{\partial \Gamma_{\alpha}^{(N)}(X,\psi)}{\partial \psi}\\
	&=& -1 + \frac{1}{[N\alpha]} \sum_{i=1}^N \theta(\psi-X_i) \nonumber\\
&\Leftrightarrow& \psi \in [X_{[ N\alpha ]: N},X_{[ N\alpha ]+1: N}) \nonumber
\end{eqnarray}
This tells us that $\psi$  splits the data sample $\{X_i\}_{i=1,\ldots,N}$ into two subsets:  $\psi$ in the minimization process moves to the specified ${\alpha}$--quantile and separates the worst $[N\alpha]$ outcomes from the remaining $N-[N\alpha]$. This also explains why the introduction of $\psi$ allows to get rid of the sorting procedure.

A single data split here is enough. Notice from eq. (\ref{estES}) that in order to estimate the expected shortfall there's no need to order all the outcomes $X_i$: it is enough to split the data sample into two subsets, and then make the average of the worst $[N\alpha]$ outcomes. A single auxiliary variable $\psi$ does the job and any finer distinction among the data is unnecessary. We will see that this is not the case when a more general spectral measure is minimized.

\section{Minimization of general Spectral Measures} \label{general}

We now turn to the problem of extending the method of Pflug--Rockafellar--Uryasev to a general Spectral Measure $M_\phi$. The hints we have obtained from studying the case of Expected Shortfall make it easier to study the extension in the discrete case first. 

\subsection{Minimization in the case of a finite number of scenarios}
The estimator of a Spectral Measure $M_\phi$ is given by
 \cite{A01}
\begin{equation} \label{estM}
M_\phi^{(N)}(X) = - \sum_{i=1}^N \phi_i X_{i:N}
\end{equation}
where $\phi_i$ is the natural discretization of $\phi$ and is itself an admissible spectrum in a discrete sense ( i.e. $\phi_i\geq 0$, $\phi_i\geq\phi_{i+1}$ $\forall i$ and $\sum \phi_i = 1$).
It is immediately clear from (\ref{estM}) that  the sorting procedure of the outcomes $X_i$ in the general case can not be replaced by a splitting into two subsets. In (\ref{estM}), all the ordered statistics $X_{i:N}$ have to be distinguished from one another since they bear in general a different weight $\phi_i$. It would be therefore perfectly useless to search for a generalization of (\ref{gammaURdiscrete}) with a single auxiliary variable $\psi$. The general solution will require $N$ auxiliary variables $\psi_i$ in order to separate all the ordered statistics from one another.

Let us define $\Delta \phi_i \equiv  \phi_{i+1}-\phi_i $ for $i=1,\ldots,N-1$ and $\Delta\phi_N\equiv -\phi_N$. We introduce a function  $\Gamma_\alpha(X,\vec{\psi})$
depending on a vector $\vec{\psi}=\{\psi_1,\psi_2,\ldots,\psi_N\}$ of auxiliary variables.
\begin{equation} \label{discretegamma}
\Gamma_\phi^{(N)}(X,\vec{\psi}) = 
\sum_{j=1}^N  \Delta\phi_j 
\left\{ 
j\,\psi_j - \sum_{i=1}^N   (\psi_j-X_i)^+ 
\right\}
\end{equation}
Let us show its extremal properties:
\begin{eqnarray} \label{extremaldiscrete}
0 	&=& \frac{\partial \Gamma_\phi^{(N)}(X,\vec{\psi})}{\partial \psi_k}\\
	&=&  \Delta\phi_k  \left[ k - \sum_{i=1}^N   \theta(\psi_k-X_i) \right] \nonumber \\
&\Leftrightarrow&
\left\{ 
\begin{array}{ll}
\psi_k = \psi_k^* \in [X_{ k : N},X_{ k+1 : N})& \mbox{if} \: \Delta\phi_k \neq 0\\
\psi_k = \mbox{whatever} & \mbox{if} \: \Delta\phi_k = 0
\end{array}
\right. \nonumber
\end{eqnarray}
Inserting the extremal condition in the functional one immediately obtains
\begin{eqnarray}
\min_{\vec{\psi}}
\left\{
\Gamma_\phi^{(N)}(X,\vec{\psi}) 
\right\}
&=& \sum_{j=1}^N \Delta\phi_j   \left[ j \,\psi_j^\star - \sum_{i=1}^N 
  (\psi_j^\star-X_i)^+  \right]\\
&=& \sum_{j=1}^N \Delta\phi_j   \left[ j \,\psi_j^\star - \sum_{i=1}^N 
  (\psi_j^\star-X_{i:N})^+  \right]\nonumber\\
&=& \sum_{j=1}^N \Delta\phi_j   \left[ j \,\psi_j^\star - \sum_{i=1}^j 
  (\psi_j^\star-X_{i:N})  \right]\nonumber\\
&=& \sum_{j=1}^N \Delta\phi_j   \left[ j \,\psi_j^\star - j 
  \psi_j^\star + \sum_{i=1}^j X_{i:N}  \right]\nonumber\\
&=& \sum_{i=1}^N X_{i:N} \sum_{j=i}^N \Delta\phi_j    \nonumber\\
&=& - \sum_{i=1}^N \phi_i X_{i:N} \nonumber \\
&=&  M_\phi^{(N)}(X)\nonumber 
\end{eqnarray}
where we have used $\sum_{j=i}^N \Delta\phi_j = -\phi_i$.

From eq. (\ref{extremaldiscrete}) it is easy to notice that $\psi_N$ plays no role in the minimization. The minimum is always achieved for $\psi_N$ large enough as far as  $\psi_N \geq {\rm ess}.\sup \{X\}$, so we can always take the limit $\psi_N\to +\infty$. We can therefore redefine $\Gamma_\phi^{(N)}(X,\vec{\psi})$ as a function of $\psi_1,\ldots,\psi_{N-1}$ only.
\begin{eqnarray}
\Gamma_\phi^{(N)}(X,\vec{\psi}) &=& 
\sum_{j=1}^{N-1}  \Delta\phi_j 
\left\{ 
j\,\psi_j - \sum_{i=1}^N  \,\, (\psi_j-X_i)^+ 
\right\}
- 
\lim_{\psi_N \to +\infty}
\phi_N 
\left\{ 
N\,\psi_N - \sum_{i=1}^N  \,\, (\psi_N-X_i)^+ 
\right\} \nonumber \\
 &=& 
\sum_{j=1}^{N-1}  \Delta\phi_j 
\left\{ 
j\,\psi_j - \sum_{i=1}^N   \,\,(\psi_j-X_i)^+ 
\right\}
- \phi_N 
\sum_{i=1}^N   X_i 
\end{eqnarray}
The function $\Gamma_\phi^{(N)}(X,\vec{\psi})$ is convex in all its parameters $(X,\vec{\psi})$ i.e. 
\begin{equation} \label{convexity}
\Gamma_\phi^{(N)}(\lambda \, X_1 + (1-\lambda)\, X_2 ,\lambda \,\vec{\psi}_1 +(1-\lambda) \,\vec{\psi}_2 ) 
\leq
\lambda\, \Gamma_\phi^{(N)}(X_1,\vec{\psi}_1) 
+ (1-\lambda)\,
\Gamma_\phi^{(N)}(X_2,\vec{\psi}_2) 
\end{equation}
for all $\lambda \in [0,1]$, provided that $\phi$ is an admissible risk spectrum. A direct check of (\ref{convexity}) is straightforward, using the convexity of $x\mapsto (x)^+$ and the decreasingness of an admissible risk aversion function which implies $\Delta\phi_i\leq 0$.

To sum up we have proved the following 

\begin{proposition} \label{prop:discrete}
Let  $M_\phi^{(N)}(X)$ be defined by eq. (\ref{estM}) and  $\Delta \phi_i \equiv  \phi_{i+1}-\phi_i $. Then, the function 
\begin{equation} \label{gammaN}
\Gamma_\phi^{(N)}(X,\vec{\psi}) =
\sum_{j=1}^{N-1}  \Delta\phi_j 
\left\{ 
j\,\psi_j - \sum_{i=1}^N \,\,  (\psi_j-X_i)^+ 
\right\}
- \phi_N 
\sum_{i=1}^N   X_i 
\end{equation}
in $N-1$ auxiliary parameters $\psi_k$ is a convex, piecewise linear funcion in all its arguments $(X,\vec{\psi})$. 
Its  minimum value  with respect to $\vec{\psi}$ equals $M_\phi^{(N)}(X)$.
\begin{equation}
\min_{\vec{\psi}} \Gamma_\phi^{(N)}(X,\vec{\psi}) = M_\phi^{(N)}(X)
\end{equation}
In the minimum, the following conditions  on $\vec{\psi}$ hold
\begin{equation}
\psi_k = \psi_k^* \in [X_{ k : N},X_{ k+1 : N}) \:\:\:\:  \mbox{if} \:\:\:\: \Delta\phi_k \neq 0
\end{equation}
while there's no dependence in  $\psi_k$  if  $\Delta\phi_k = 0$.
\end{proposition}

Notice that the piecewise--linearity of the function $\Gamma_\phi^{(N)}$ of (\ref{gammaN}) allows for a linearization of its minimization problem following the very same technique introduced in \cite{Pf00,RU00,RU01}. We will exploit this fact in a forthcoming publication which will analyze concrete applications \cite{AGNSV}.

\subsection{Minimization in the general case}

It is now easy to obtain an analogous functional for the continuous case of a Spectral Measure $M_\phi$ of eq. (\ref{Mphi}). 
The continuous limit of eq. (\ref{gammaN}) is straightforward
\begin{equation}\label{contgamma}
\Gamma_\phi[\psi] = 
\int_0^1 d\alpha\: \frac{d\phi}{d\alpha}
\left\{ 
\alpha
\psi(\alpha) -  \mathrm{E}[\psi(\alpha)-X]^+
\right\}  - 
\phi(1) \,
 \mathrm{E}[X]
\end{equation}
The function $\psi$ on which  we take variations is defined on the open interval $(0,1)$ only. The extremal condition gives 
\begin{eqnarray}
0 	&=& \frac{\delta \Gamma_\phi[\psi] }{\delta \psi(\beta)}\\
	&=& 
\int_0^1 d\alpha\: \frac{d\phi}{d\alpha}
\left\{ 
\alpha\,
\delta(\alpha-\beta) -  \mathrm{E}[\theta(\psi(\alpha)-X)]\,\delta(\alpha-\beta)
\right\}
 \nonumber \\
	&=& 
\int_0^1 d\alpha\: \frac{d\phi}{d\alpha}
\left\{ 
\alpha
 -  F_X(\psi(\alpha))
\right\}\delta(\alpha-\beta)
 \nonumber \\
&\Leftrightarrow& 
\left\{
\begin{array}{ll}
\psi(\beta) =\psi^\star(\beta) \in [ F^\leftarrow_X (\beta),F^\rightarrow_X (\beta) ) & \mbox{if} \:\: \displaystyle \frac{d\phi}{d\beta} 
\neq 0 \\\\
\mbox{no conditions on }\psi(\beta)& \mbox{if} \:\: \displaystyle \frac{d\phi}{d\beta}= 0 
\end{array}
\right.
\nonumber
\end{eqnarray}
Inserting $\psi^*$ into eq. (\ref{contgamma}) we obtain
\begin{eqnarray}
\min_{\psi}
\left\{
\Gamma_\phi[\psi]
\right\}
&=& \Gamma_\phi[\psi^\star]  \\
&=& 
\int_0^1 d\alpha\: \frac{d\phi}{d\alpha}
\left\{ 
\alpha
F_X^\leftarrow(\alpha) -  \int_0^1 dp \:(F_X^\leftarrow(\alpha)-F_X^\leftarrow(p))^+
\right\}
- 
\phi(1) \,
 \mathrm{E}[X]
\nonumber \\
&=& 
\int_0^1 d\alpha\: \frac{d\phi}{d\alpha}
\left\{ 
\alpha
F_X^\leftarrow(\alpha) -  \int_0^\alpha dp \:(F_X^\leftarrow(\alpha)-F_X^\leftarrow(p))
\right\}
- 
\phi(1) \,
 \mathrm{E}[X]
\nonumber \\
&=& 
\int_0^1 d\alpha\: \frac{d\phi}{d\alpha}
  \int_0^\alpha dp \:F_X^\leftarrow(p)
- 
\phi(1) \,
 \mathrm{E}[X]
\nonumber \\
&=& 
-\int_0^1 d\alpha\: \phi(\alpha)
  F_X^\leftarrow(\alpha)
+ 
\phi(1) \,
 \mathrm{E}[X]
- 
\phi(1) \,
 \mathrm{E}[X]
\nonumber \\
&=&
M_\phi \nonumber
\end{eqnarray}

Summarizing, we have proved the following 

\begin{proposition} \label{prop:cont}
Let  $M_\phi(X)$ be defined by eq. (\ref{Mphi}). Then, the functional 
\begin{equation} \label{gamma}
\Gamma_\phi[X,\psi] = 
\int_0^1 d\alpha\: \frac{d\phi}{d\alpha}
\left\{ 
\alpha
\psi(\alpha) -  \mathrm{E}[\psi(\alpha)-X]^+
\right\}  - 
\phi(1) \,
 \mathrm{E}[X]
\end{equation}
in the function $\psi:(0,1)\to\mathrm{R}$ is a convex, piecewise linear functional in all its arguments $[X,{\psi}]$. 
Its  minimum value  with respect to ${\psi}$ equals $M_\phi(X)$.
\begin{equation}
\min_{{\psi}} \Gamma_\phi[X,{\psi}] = M_\phi(X)
\end{equation}
In the minimum, the following conditions  on ${\psi}$ hold
\begin{equation}
\psi(\alpha) =\psi^\star(\alpha) \in [ F^\leftarrow_X (\alpha),F^\rightarrow_X (\alpha) ) \:\:\:\:\:\: \mbox{if} \:\:\:\:\:\:  \frac{d\phi}{d\alpha}\neq 0 \end{equation}
while there's no dependence on  $\psi(\alpha)$  if  $\frac{d\phi}{d\alpha} = 0$.
\end{proposition}

\begin{remark} \label{rmk}\rm
Consider the case of a piecewise--constant admissible risk spectrum $\phi$ with $J$ jumps $\Delta\phi_k= \phi(p_k^+)-\phi(p_k^-)$ $k=1,\ldots,J$ at $p=p_1,\ldots,p_J <1$. Observing that 
\begin{equation}
\frac{d\phi}{d\alpha} =  \sum_{k=1}^J \Delta\phi_k \delta(\alpha-p_k)
\end{equation}
a straightforward application of Proposition \ref{prop:cont} shows that the functional $\Gamma_\phi$ reduces to a function in $J$ auxiliary variables only $\psi_k\equiv\psi(p_k)$ 
\begin{equation} \label{piecewise}
\Gamma_\phi[X,\psi]=\Gamma_\phi(X,\vec{\psi}) = 
\sum_{k=1}^J \Delta\phi_k
\left\{ 
p_k
\psi_k -  \mathrm{E}[\psi_k-X]^+
\right\}  - 
\phi(1) \,
 \mathrm{E}[X]
\end{equation}
\end{remark}

\section{Maximizing returns and minimizing risks altogether: \\connection with the Markowitz risk--reward approach}

Assume now that the p\&l r.v. $X=X(\vec{w})$ depends on a set of $W$ parameters $w_k$ and let ${\cal W}\subset \mathbb{R}^W$ be a set of acceptable weights. Exploiting the result of Proposition \ref{prop:cont}, the problem of minimization of  a specified Spectral Measure $M_\phi(X(\vec{w}))$ with  constraints $\vec{w}\in {\cal W}$ can be mapped into the equivalent  minimization problem of the functional $\Gamma_\phi[X(\vec{w}),\psi]$
\begin{equation}
\left|
\begin{array}{l}
\displaystyle
\min_{\vec{w}} M_\phi(X(\vec{w})) \\\\
\displaystyle
\vec{w} \in \cal{W}
\end{array}
\right.
\hspace{1cm}
\longleftrightarrow
\hspace{1cm}
\left|
\begin{array}{l}
\displaystyle
\min_{\vec{w},\psi} \Gamma_\phi[X(\vec{w}),\psi] \\\\
\displaystyle
\vec{w} \in \cal{W}
\end{array}
\right.
\end{equation}
It is interesting to study the problem of optimization of a portfolio via simultaneous minimization of a specified spectral measure $M_\phi(X)$ and maximization of the expected return $\mathrm{E}[X]$ of the portfolio. In other words, we want to study the problem of determining the efficient frontier on the Markowitz plane $(x,y)\equiv(M_\phi(X),\mathrm{E}[X])$.

Let us now extend some standard terminology for the $(M_\phi(X),\mathrm{E}[X])$ risk--reward optimization problem:

\begin{definition} \label{optim}
In the  $(M_\phi(X),\mathrm{E}[X])$ risk--reward optimization problem with domain ${\cal W }$, we will say that a portfolio $\pi(\vec{z})$ {\bf dominates}  $\pi(\vec{w})$  (and we will write $\pi(\vec{z})\succ\pi(\vec{w})$) if $\vec{w},\vec{z}\in{\cal W}$ and
\begin{equation}
\mathrm{E}[X(\vec{z})] \geq \mathrm{E}[X(\vec{w})] 
\hspace{1cm} \mbox{and} \hspace{1cm} 
M_\phi(X(\vec{z})) \leq M_\phi(X(\vec{w})) 
\end{equation}
where at least one of the two inequalities has to be strict. 

We will say that a portfolio $\pi(\vec{w})$ is {\bf optimal} if there is no $\vec{z}\in{\cal W}$ such that  $\pi(\vec{z})\succ\pi(\vec{w})$. The geometrical set of all optimal portfolios will be called the {\bf efficient frontier} of the plane $(M_\phi(X),\mathrm{E}[X])$.

\end{definition}

The $(M_\phi(X),\mathrm{E}[X])$ optimization problem can be naturally set up as a constrained problem where 
$M_\phi$ is minimized for a specified value of return $\mathrm{E}[X(\vec{w})] = \mu$
\begin{equation} \label{prob:constr1}
\left|
\begin{array}{l}
\displaystyle
\min_{\vec{w}} M_\phi(X(\vec{w})) \\\\
\displaystyle
\mathrm{E}[X(\vec{w})] = \mu \\\\
\displaystyle
\vec{w} \in \cal{W}
\end{array}
\right.
\end{equation}
or alternatively in the specular problem in which the expected return is maximized for a specified value $M_\phi(X(\vec{w})) = \rho$ of risk 
\begin{equation}\label{prob:constr2}
\left|
\begin{array}{l}
\displaystyle
\max_{\vec{w}} \mathrm{E}[X(\vec{w})] \\\\
\displaystyle
M_\phi(X(\vec{w})) = \rho \\\\
\displaystyle
\vec{w} \in \cal{W}
\end{array}
\right.
\end{equation}
Both problems can be faced replacing  $M_\phi(X(\vec{w}))$ with the functional $\Gamma_\phi[X(\vec{w}),\psi]$. Pflug, Rockafellar and Uryasev \cite{Pf00,RU00,RU01} extensively study these optimization problems in the particular case of Expected Shortfall showing that they may be reduced to equivalent problems solvable by linear programming. 

Remember that in order to solve both problems (\ref{prob:constr1}) and (\ref{prob:constr2})  a final check is necessary to make sure that the minimizing (resp. maximizing) portfolio is in fact an optimal portfolio in the sense of Definition \ref{optim}. In other words, the chosen values of the constraints $\mu$ or $\rho$ may turn out to be not compatible with the efficient frontier. Only studying the global form of the frontier one can assess the optimality of solutions. 

In this paper we adopt a slightly different approach from constrained optimization as in (\ref{prob:constr1}) and (\ref{prob:constr2}). We will in fact exploit the fact that minimizing a Spectral Measure is already in some sense {\em ``minimizing risks and maximizing returns at the same time''}. A spectral risk measure is a weighted average of all possible profits and losses of a portfolio. The risk aversion function $\phi$ in the general case when $\phi(1)>0$ takes into account all possible future scenarios including the extreme returns. In the minimization of  a spectral measure it is impossible to disentangle minimization of risks from maximization of returns.

We can now show that the $(M_\phi(X),\mathrm{E}[X])$ risk--reward optimization problem is equivalent to the minimization problem of a suitable family of different spectral measures $M_{\hat{\phi}(\lambda)}$.

\begin{proposition} \label{prop:optim} Let $M_\phi$ be a spectral risk measure. The optimal portfolios of the corresponding  \linebreak $(M_\phi(X),\mathrm{E}[X])$ risk--reward optimization problem are the solutions of the (unconstrained) minimization problem of the spectral measures 
\begin{equation} \label{minprob}
M_{\hat{\phi}(\lambda)} (X) = - \lambda \, \mathrm{E}[X] + (1-\lambda) \, M_\phi (X)
\end{equation}
defined for $\lambda \in [0,1]$.
\end{proposition}
{\bf proof:} first of all, to check that $M_{\hat{\phi}(\lambda)}$ is indeed a spectral measure for $\lambda\in [0,1]$, we observe that   $-\mathrm{E}[X] = ES_{1}(X)$ is a spectral measure and that a convex combination of spectral measures is a spectral measure. It is easy to verify that its risk aversion function $\hat{\phi}(\lambda)=  \lambda  + (1-\lambda) \, \phi$ is an admissible risk spectrum.

We can use a Lagrange multiplier $\beta$ to solve the problem of minimization of $M_\phi$ with contraints on the return $\mathrm{E}[X]=\mu$.
\begin{equation}\label{eq:lagrange1}
\left\{
\begin{array}{l}
\Delta_X M_\phi (X) -\beta  \Delta_X \mathrm{E}[X] = 0 \\
\mathrm{E}[X] = \mu
\end{array}
\right.
\end{equation}
where we have denoted as $\Delta_X$ the differential\footnote{For arbitrary unconstrained changes $\Delta\vec{w}$  the differential operator is explicitly written as
$$
\Delta_X = \sum_k \, \Delta w_k \, \frac{\partial}{\partial w_k} 
$$} with respect to arbitrary changes in $X$ induced by possible changes $\Delta\vec{w}$.
Solving the first equation with respect to the weights $\vec{w}$ one obtains a solution $\vec{w}=\vec{w}(\beta)$ for any value of the Lagrange multiplier. Then, the solution will be selected by imposing $\mathrm{E}[X(\beta)] = \mu$.
From Definition \ref{optim} it is clear that the solution will be an optimal portfolio if and only if $\beta>0$, otherwise it would be possible a small change in the parameters $\vec{w}$ such that  $\Delta M_\phi \leq 0$ and $\Delta \mathrm{E}[X]\geq 0$. The minimization problem (\ref{minprob}) is defined by the equation
\begin{equation} \label{eq:lagrange2}
(1-\lambda) \, \Delta_X M_\phi (X) -\lambda  \Delta_X \mathrm{E}[X] = 0 
\end{equation}
It is clear that the solutions of eq. (\ref{eq:lagrange2}) coincide with those of the problem (\ref{eq:lagrange1}) if $\beta = \lambda/(1-\lambda)$. It is also clear that the solutions of (\ref{eq:lagrange2}) will automatically select only the optimal solutions, since for $\lambda \in [0,1]$, we have $\beta\in [0,\infty)$. \hfill $\clubsuit$.

\vspace{0.5cm}
Proposition \ref{prop:optim} shows that the constrained optimization problems (\ref{prob:constr1}) and (\ref{prob:constr2}) are in fact equivalent to a one--parameter family of non--constrained minimization problems in a single spectral measure $M_{\hat{\phi}(\lambda)} (X)$ which linearly interpolates between $M_\phi$ and $-\mathrm{E}[X]$. 
\begin{remark} \rm
Notice in particular that the minimization of $M_{\hat{\phi}(\lambda)} (X)$  automatically selects only optimal portfolios in the sense of Definition \ref{optim} and there is no need for an eventual check. The above proof shows that the non--optimal solutions of the constrained  problems (\ref{prob:constr1}) and (\ref{prob:constr2}) correspond to minima of $M_{\hat{\phi}(\lambda)} (X)$ for $\lambda \notin [0,1]$ i.e. minima of a measure of risk which in general is not coherent.
\end{remark}

\begin{example} \rm
As the simplest example one can for instance study the case of the optimization problem in the plane $(ES_\alpha(X),\mathrm{E}[X])$ for some specified quantile level $\alpha$. The efficient frontier can be obtained by minimizing $\forall \lambda \in [0,1]$ the spectral measure defined by
\begin{equation} 
M_{\hat{\phi}(\lambda)} (X) = - \lambda \, \mathrm{E}[X] + (1-\lambda) \, ES_\alpha (X)
\end{equation}
with risk aversion function 
\begin{equation}
\hat{\phi}(\lambda)(p) =  \lambda  + \frac{1-\lambda}{\alpha} \, \theta(\alpha - p)
\end{equation}
This risk aversion function is piecewise--constant $\forall \lambda$ and has a single jump $\Delta\hat{\phi}= (1-\lambda)/\alpha$ in the interval $p\in(0,1)$. Hence, using the observation of Remark \ref{rmk}, this minimization problem can be solved by adopting the function $\Gamma_{\hat{\phi}(\lambda)}(X,\psi_1)$  of eq. (\ref{piecewise}) with a single auxiliary variable $\psi_1$, $p_1=\alpha$, $\hat{\phi}(p=1)= \lambda$ and $\Delta\hat{\phi}_1= (1-\lambda)/\alpha$.
\end{example}

\section{Conclusions}

In this paper we have derived the extension to general spectral measures of the results of Pflug--Rockafellar-Uryasev on the optimization of Expected Shortfall. The complexity of the optimization of a spectral measure is dramatically reduced by showing that it coincides to the optimization problem of a suitable alternative function which is convex and piecewise--linear in the portfolio weights plus a number of additional parameters (Propositions \ref{prop:discrete} and \ref{prop:cont}). This result permits to set up efficient optimization algorithms for the minimization problem of a general spectral measure. The problem can be reduced to simple linear programming \cite{AGNSV}.

The Markowitz problem of finding the optimal portfolios minimizing a spectral measure $M_\phi(X)$ and maximizing returns $E[X]$ is shown to be equivalent to the unconstrained minimization problem of minimizing a suitable spectral measure which interpolates $M_\phi(X)$ and $-E[X]$ (Proposition \ref{prop:optim}). This result makes it clear that the minimization of a spectral measure already takes into account risk minimization and returns maximization.

In a problem \`a la Markowitz, the investor expresses her risk aversion only through a single constraint which specifies a chosen level of risk or return. In the realm of spectral measures  an investor can optimize a portfolio in a more articulated way by expressing her subjective risk aversion via the function $\phi$.  Splitting this process in two (risk minimization and returns maximization) turns out to be impossible and somewhat artifical.


\newpage
\sloppy

\end{document}